\documentclass[preprint,aps,showpacs,prb]{revtex4}

\usepackage{graphicx}
\usepackage{dcolumn}
\usepackage{bm}

\nonstopmode
\usepackage{epsfig}

\graphicspath{{./}{data_paper/}}
\DeclareGraphicsExtensions{.ps,.ps.gz}
\begin{document}

\title{\bf On a method to calculate conductance by means of the Wigner function: two critical tests}

\author{Ioan B\^aldea}
\email{ioan.baldea@pci.uni-heidelberg.de}
\altaffiliation[Also at ]{National Institute for Lasers, Plasma, 
and Radiation Physics, ISS, POB MG-23, RO 077125 Bucharest, Romania.}
\author{Horst K\"oppel} 
\affiliation{Theoretische Chemie,
Physikalisch-Chemisches Institut, Universit\"{a}t Heidelberg, Im
Neuenheimer Feld 229, D-69120 Heidelberg, Germany}

\date{\today}

\begin{abstract}
We have implemented the linear response approximation of a method proposed to compute the 
electron transport through correlated molecules 
based on the time-independent Wigner function 
[P.\ Delaney and J.\ C.\ Greer, \prl {\bf 93}, 036805 (2004)]. 
The results thus obtained for the zero-bias conductance through a quantum dot both without 
and with correlations demonstrate that this method is neither quantitatively nor qualitatively 
able to provide a correct physical description of the electric transport through nanosystems. 
We present an analysis indicating that the failure is due to the manner of imposing the 
boundary conditions, and that it cannot be simply remedied.
\end{abstract}




\maketitle

\renewcommand{\topfraction}{1}
\renewcommand{\bottomfraction}{1}
\renewcommand{\textfraction}{0}
\section{Introduction}
\label{sec:introduction}
Electronic transport in artificial nanosystems and single molecules represents a 
topic of continuing current interest and remains a challenge 
both for fundamental science and technological applications.
In spite of numerous advances, there still exist important issues which  
are far from being resolved.
Notoriously, the comparison between experimental and theoretical values for the 
current flowing through molecules 
usually show large discrepancies, typically several orders of 
magnitudes.\cite{Wenzel:02,Nitzan:03,Stokbro:03}
No consensus has been reached so far whether the discrepancies are to be attributed to 
uncontrollable experimental factors (compare measurements on same systems in Refs.\
\onlinecite{Reed:97,Xiao:04,Daosh:05,tsutsui:06}) 
or inadequate theoretical frameworks.
\par
Early theoretical calculations were carried out within the 
Landauer formalism,\cite{Landauer:70,Buttiker:85} which is based on  
a single-particle description. Because systematically the values of the 
currents thus obtained are at odds with those experimentally measured, 
a series of theoretical methods was developed 
to treat electron correlations, which are excluded by the Landauer approach.
\par
The description of electron correlation represents an important challenge 
from the theoretical side.
The most popular theoretical approaches for nonequilibrium transport 
in correlated nanoscopic/molecular systems are based on nonequilibrium 
Green functions (NEGF),\cite{HaugJauho,Datta:05} time-dependent DMRG (density matrix 
renormalization group), \cite{White:04,Daley:04,White:05} and  
numerical renormalization group (NRG).\cite{Bulla:08}
In \emph{ab initio} modeling of molecular electronics, the NEGF technique 
is by far the most utilized one. 
It is usually combined with electron structure calculations based on 
density functional theories (DFT).\cite{Taylor:01,Brandbyge:02,Rocha:06} 
The NEGF + DFT approach is conceptually simple and computationally less demanding than 
many-body methods.\cite{Wenzel:03,DelaneyGreer:04a,DelaneyGreer:04b,Muralidharan:06,Thygesen:07}
However, because of the incompletely elucidated current dependence \cite{Vignale:87,Vignale:88} 
and self-interaction corrections \cite{toher:07,Ke:07} 
of the local exchange and correlation functionals,
the DFT currents can sensitively vary by choosing different approximate potentials.
\par
Out of the methods proposed so far, 
the method proposed in Ref.\ \onlinecite{DelaneyGreer:04a} seems to be one of the most appealing, 
because of its claim to correctly reproduce the steady-state current experimentally measured 
through correlated molecular systems.
It relies upon genuine many-body calculations and is not limited to a particular 
configuration interaction (CI), \emph{e.\ g.,} restricted to singly, doubly, or triply-excited 
configurations. The originality of this approach consists in the manner 
of imposing the boundary conditions. Rather than the Fermi distribution function, 
the Wigner distribution function is employed instead for the constrains at the device-electrodes 
(reservoirs) interfaces. 
Unlike the former, which is meaningful only within a single-particle picture, the latter can be directly 
used in a many-body approach.
Because the key quantity of this method is the stationary 
Wigner function (SWF), hereafter it will be referred to as the SWF-method. 
\par
One should emphasize 
from the very beginning that the support of the SWF-method, and especially its manner 
to impose boundary conditions, entirely relies upon its ability to provide 
currents comparable 
with those measured experimentally for two molecular systems \cite{DelaneyGreer:04a,Fagas:07} 
and plausible results for another class of molecules.\cite{Fagas:06}
Because this method is conceptually so different from the other widely utilized approaches, 
it would be highly desirable to inquire the validity of this method within the realm of theory.
If and to the extent to which the SFW-method is able to reproduce well established 
results, one can consider its predictions reliable. 
It is the purpose of the present work to investigate whether the SWF-method is able 
to correctly describe nanosystems whose behavior is well understood.
\par
The remaining part of the paper is organized in the following manner. 
The theoretical framework, the linear response approximation of the SWF-method, 
will be developed 
in Sec.\ \ref{sec:method}. The model for the ``device'' we shall consider, 
consisting of a single quantum dot (QD), will be exposed in Sec.\ \ref{sec:model}.
Next, in Sec.\ \ref{sec:uncorrelated}, we shall compare the zero-bias 
conductance calculated without correlations by means of the SWF-method with 
the exact one. Sec.\ \ref{sec:correlated} will be devoted to the linear 
transport through the QD in the presence of correlations. 
The implications of our results will be discussed in Sec.\ \ref{sec:discussion}.
Conclusions will be presented in Sec.\ \ref{sec:conclusion}.
\section{Method}
\label{sec:method}
Theoretical studies of the transport in nanosystems are inherently 
faced with the problem of separating the total system ($\mathcal{S}$) into 
left and right reservoirs (electrodes), 
and device. The contacts (interfaces) between device and reservoirs 
(located at $q_{L,R}$) are necessarily subject to arbitrariness. 
Depending on how demanding the numerical calculations are, 
parts of the electrodes (as large as possible) are included in the central part 
and treated as accurate as possible. In the SFW-method, the challenge is 
to determine the wave function $\Psi$ for the total system $\mathcal{S}$.
\par
Following Ref.\ \onlinecite{DelaneyGreer:04a}, we shall consider the 
many-electron wave function $\Psi$ for the total system $\mathcal{S}$ 
and its associate exact energy $\mathcal{E}$ in the presence of an applied bias
\begin{equation}  
\label{eq-total-energy}  
\mathcal{E} = \left \langle \Psi \right \vert H_{T} \left \vert \Psi \right \rangle 
= \left \langle \Psi \right \vert H \left \vert \Psi \right \rangle
+ \left \langle \Psi \right \vert W \left \vert \Psi \right \rangle ,
\end{equation}  
where $H$ is the Hamiltonian of the total system $\mathcal{S}$, 
and $W$ is the perturbation due to applied bias.
\par
Within the single-particle picture, in a transport problem the (semi-infinite) 
reservoirs fix the electron chemical potentials $\mu_{L, R}$ 
at the left and right contacts, 
and the current flow is driven by the imbalance $\mu_L - \mu_R = e V$
created by an applied voltage. At either contact, the electron distribution 
is dictated by the reservoir Fermi function with the correponding chemical potential.
The attractive feature of the approach proposed in 
Ref.\ \onlinecite{DelaneyGreer:04a} is that it is based on the Wigner function,
which is directly applicable to a many-body system, without the need 
to resort to single-particle Fermi distribution functions.
For problems describable within a single-particle picture, 
the Wigner function method to account for open boundaries specific for 
transport problems has certain advantages over the conventional scattering 
approach.\cite{Frensley:91,Frensley:86,Frensley:88}
However, these advantages are merely technical there.
\par
For a system characterized by a many-electron wave function $\Psi$, 
the one-particle Wigner distribution function $f(q, p)$ is defined by
\begin{equation}
f(q, p) = \sum_{r, \sigma} \left\langle \Psi \right \vert 
a^{\dagger}_{q-r,\sigma} a_{q+r,\sigma} \left \vert \Psi \right \rangle
e^{-2 i p r/\hbar} .
\label{eq-wigner}  
\end{equation}  
Unlike the methods based on the NEGF, where semi-infinite leads are accounted for,
the SWF-method was devised to be suited for \emph{ab initio} 
quantum chemical calculations, where rather than semi-infinite electrodes, 
only (usually very) small parts thereof can be dealt with.
Instead of having an infinite extension, the wave function $\Psi$ 
and the related summation over $r$ of Eq.\ (\ref{eq-wigner}) 
are inherently limited in space.
In this paper, we shall adopt a one-dimensional discrete representation, wherein 
in a lattice of size $N=2M+1$ the site index ranges from $-M$ to $+M$. 
In Eq.\ (\ref{eq-wigner}), $a_{l,\sigma}$ ($a^{\dagger}_{l,\sigma}$) denote the creation   
(annihilation) operators for electrons of spin $\sigma$ at site $l$.
\par
The open boundary conditions to be imposed should distinguish between 
electrons flowing into and those flowing out of the device.\cite{Frensley:91} 
The distribution of 
the former should be determined by the reservoirs. In terms of the Wigner function, 
this means that, 
at the left ($q_L$) and right ($q_R$) boundaries between device and reservoirs,  
$f\left(q_{L}, p_{L} > 0\right)$ and $f\left(q_{R}, p_{R} < 0\right)$ should be fixed 
at values dictated by the reservoirs \cite{Frensley:91,Frensley:86,Frensley:88} 
\begin{eqnarray}  
\label{eq-FW-LB}  
& & f\left(q_{L}, p_{L}\right) = f_{L}\left(p_{L}; \mu_{L}\right), \mbox{ for } p_{L}>0, \\
\label{eq-FW-RB}  
& & f\left(q_{R}, p_{R}\right) = f_{R}\left(p_{R}; \mu_{R}\right), \mbox{ for } p_{R}<0 .
\end{eqnarray}  
These constraints are physically plausible and have 
been successfully applied previously to problems treated within single particle 
approximations.\cite{Frensley:91,Frensley:86,Frensley:88} The 
r.h.s.\ of Eqs.\ (\ref{eq-FW-LB}) and (\ref{eq-FW-RB}) represent the Fermi functions of 
the reservoirs with the chemical potentials shifted by the bias voltage,
$\mu_{L,R} = \mu \pm eV/2$.
Because this procedure cannot be applied for the many-body case, to determine
the above $f_{L,R}$, which are basically properties of the reservoirs, 
it was claimed \cite{DelaneyGreer:04a} that they can be extracted from the 
wave function of the total system $\mathcal{S}$ at \emph{zero} bias.
That is, at zero temperature, one has to determine the ground state $\Psi_{0}$
of $H$, then calculate $f_0$ by using $\Psi_{0}$ instead of $\Psi$ 
in Eq.\ (\ref{eq-wigner}), and impose
\begin{eqnarray}  
\label{eq-FW-L}  
& & f\left(q_{L}, p_{L}\right) = f_{0}\left(q_{L}, p_{L}\right), \mbox{ for } p_{L}>0, \\
\label{eq-FW-R}  
& & f\left(q_{R}, p_{R}\right) = f_{0}\left(q_{R}, p_{R}\right), \mbox{ for } p_{R}<0 .
\end{eqnarray}  
\par
To describe a steady state, 
one has to impose in addition the condition that the 
average $J$ of the electric current operator ($e$ is the elementary charge)\cite{Caroli:71}
\begin{equation}
\label{eq-j}  
j_{l} = i t_{l} \frac{e}{\hbar}\sum_{\sigma} \left(
a^{\dagger}_{l+1,\sigma} a_{l,\sigma} - a^{\dagger}_{l,\sigma} a_{l+1,\sigma} 
\right)
\end{equation}
be site ($l$) independent
\begin{equation}
\label{eq-j-average}    
J_{l} = \left\langle \Psi \right \vert j_{l}\left \vert \Psi \right \rangle = J.
\end{equation}  
Above, $t_{l}$ denotes the hopping integral ($-M \leq l \leq M - 1$). 
\par
According to Ref.\ \onlinecite{DelaneyGreer:04a}, the solution 
of the transport problem is obtained by minimizing $\mathcal{E}$ with 
the supplementary constraints 
(\ref{eq-FW-L}), (\ref{eq-FW-R}), and (\ref{eq-j-average}) along with the 
normalization condition 
\begin{equation}  
\label{eq-norm-Psi}
\left\langle \Psi \right. \left \vert \Psi \right \rangle = 1.
\end{equation} 
In this way, to reach its steady state,
the device is allowed to optimize the Wigner distribution function of 
the electrons inside the device and that of the electrons flowing from it  
into reservoirs. (The distribution of 
outgoing electrons should depend only of the state of the device.)
\par
In the present Section, we shall work out the linear response approximation
of the SFW-method, which enables us later to compute the zero-bias conductance.
That is, we shall only consider changes to the relevant 
quantities of the order $\mathcal{O}(V)$, caused by a small applied voltage $V$. 
The system (left reservoir, device, 
and right reservoir), which consists of a collection of discrete sites ranging from $-M$ to $+M$,
is perturbed by a small electric perturbation
\begin{equation}  
W = - e \sum_{l=-M}^{M} n_{l} \varphi_{l} ,
\label{eq-W}  
\end{equation}  
where $n_{l} = \sum_{\sigma}a^{\dagger}_{l,\sigma} a_{l,\sigma}$ is the 
electron number operator and $\varphi_{l}$ is the electric potential at site $l$.
($\varphi_{-M} = +V/2$, $\varphi_{+M} = -V/2$).
Starting with a system in the ground state $\Psi_{0}$, we shall look for a solution
$\Psi$ describing a steady state that slightly differs from $\Psi_{0}$. 
The wave function $\Psi$ will be expanded in terms of the complete set 
of eigenstates of $H$ ($H\vert\Psi_{n}\rangle = E_{n}\vert\Psi_{n}\rangle$)
\begin{equation}  
\left\vert \Psi \right \rangle 
=  A_{0} \left \vert \Psi_{0} \right \rangle + 
\sum_{n\neq 0} A_{n} \left \vert \Psi_{n} \right \rangle ,
\label{eq-Psi}  
\end{equation}  
where $A_{n} = \mathcal{O}(V)$ for $n\neq 0$. 
One should note at this point that, although the eigenstates $\Psi_{n}$ can and 
will be chosen real, in order to satisfy 
the boundary conditions (\ref{eq-FW-L}) and (\ref{eq-FW-R}) 
the expansion coefficients 
$A_{n}$ must be complex. While in general the minimization of 
$\mathcal{E}$, Eq.\ (\ref{eq-total-energy}), with the constraints 
(\ref{eq-FW-L}), (\ref{eq-FW-R}), (\ref{eq-j-average}), and (\ref{eq-norm-Psi}) represents a 
nonlinear problem, it considerably simplifies in the linear response limit.
By introducing the Lagrange multipliers $\lambda_{L,p_{L}}$ ($p_{L}>0$), 
$\lambda_{R,p_{R}}$ ($p_{R}<0$), $\omega$, and $\chi_{l}$ for the constraints 
(\ref{eq-FW-L}), (\ref{eq-FW-R}), (\ref{eq-norm-Psi}), and (\ref{eq-j-average}), respectively, 
for small $V$ the minimization amounts to solve a linear system of equations, 
which possesses a solution $A_{0} = 1 + \mathcal{O}(V^2)$, 
$A_{n} = \mathcal{O}(V)$ for $n\neq 0$, $\lambda_{L,p} = \mathcal{O}(V)$,
$\lambda_{R,p} = \mathcal{O}(V)$, $\omega = E_{0} + \mathcal{O}(V)$, 
and $\chi_{l} = \mathcal{O}(V)$. 
\par
The quantites entering the minimization problem within the linear response approximations are
\begin{eqnarray}
\label{eq-F}  
& & \mathcal{F}_{n}\left(q, p\right) \equiv 
\sum_{r, \sigma} \left\langle \Psi_{n} \right \vert 
a^{\dagger}_{q-r,\sigma} a_{q+r,\sigma} \left \vert \Psi_{0} \right \rangle
e^{-2 i p r/\hbar}, \\
\label{eq-J}  
& & \mathcal{J}_{n}(l) \equiv \left\langle \Psi_n \right \vert j_{l}\left \vert \Psi_0 \right \rangle , \\
\label{eq-P}  
& & \mathcal{W}_{n} \equiv \left\langle \Psi_n \left \vert W \right \vert \Psi_0 \right \rangle .
\end{eqnarray}  
For the ground state ($n=0$), $\mathcal{F}_{0}\left(q, p\right) = f_{0}(q, p)$.
\par
We shall assume (a fact justified for the models considered here) that 
in the absence of perturbation the Hamiltonian of the system $H$ is invariant under inversion, 
and the ground state $\Psi_{0}$ is even and carries no current, $\mathcal{J}_{0}(l) = 0$.
Owing to the spatial inversion of $H$, the eigenstates 
$\Psi_{n}$ are either even ($\Psi_{g}$) or odd ($\Psi_{u}$). 
In view of the space inversion, 
is it natural to choose boundaries located symmetrically 
($q_L = - q_R$), a mesh comprising symmetric values 
of positive and negative values of $p$ ($p_{L} = - p_{R} = p > 0$),
and an antisymmetric applied potential profile ($\varphi_{-l} = - \varphi_{l}$).
The quantities (\ref{eq-F}), (\ref{eq-J}), and (\ref{eq-P}) 
possess the following symmetry properties:
\begin{eqnarray}
\label{eq-sym-F}  
& & \mathcal{F}_{g}\left(q_L, p\right) = \mathcal{F}_{g}\left(q_R, -p\right), 
\mathcal{F}_{u}\left(q_L, p\right) = - \mathcal{F}_{u}\left(q_R, -p\right) , \\
\label{eq-sym-J}  
& & \mathcal{J}_{g}(l) = - \mathcal{J}_{g}(-l), \mathcal{J}_{u}(l) = \mathcal{J}_{u}(-l) , \\
\label{eq-sym-P}  
& & \mathcal{W}_{g} = 0.
\end{eqnarray}  
Straightforward algebra shows that the equations for 
even ($g$) and odd ($u$) eigenstates decouple, and only odd ($u$) 
eigenstates contribute to the current
\begin{equation}  
J = J_{l} = 2 \sum_{u} \mbox{Im} 
\mathcal{J}_{u}(l)
\mbox{Im}  A_{u} .
\label{eq-J-lra}  
\end{equation}  
The minimization yields the following linear equations for the
Lagrange multipliers 
$\lambda(p) \equiv \left[ \lambda_{L}(p) - \lambda_{R}(-p)\right]/2$
and $\chi(j) \equiv \left[ \chi(j) - \chi(-j)\right]/2$
($j, j^\prime, p, p^\prime > 0$)
\begin{eqnarray}  
\displaystyle
& & \sum_{p^{\prime}} \lambda\left( p^{\prime}\right) \sum_{u} \frac{1}{E_{u} - E_{0}} 
\left[ 
\mathcal{F}_{u}\left(q_{L}, p^\prime\right) \mathcal{F}^{\ast}_{u}\left(q_{L}, p\right) + 
\mathcal{F}^{\ast}_{u}\left(q_{L}, p^\prime\right) \mathcal{F}_{u}\left(q_{L}, p\right)
\right] \nonumber \\
\label{eq-Lagrange-FW}  
& & + \sum_{j^{\prime}=1}^{M-1} \chi(j^{\prime}) \sum_{u} \frac{1}{E_{u} - E_{0}} 
\left[ 
\mathcal{I}_{u} \left(j^{\prime}\right) \mathcal{F}^{\ast}_{u}\left(q_{L}, p\right) + 
\mathcal{I}^{\ast}_{u} \left(j^{\prime}\right) \mathcal{F}_{u}\left(q_{L}, p\right)
\right] \\
& & = \frac{1}{2}\sum_{u} \frac{\mathcal{W}_{u}}{E_{u} - E_{0}} 
\left[ 
\mathcal{F}^{\ast}_{u}\left(q_{L}, p\right) + \mathcal{F}_{u}\left(q_{L}, p\right)
\right] . \nonumber
\end{eqnarray}  
\begin{equation}  
\displaystyle
\sum_{p^{\prime}} \lambda\left( p^{\prime}\right) \sum_{u} 
\frac{\mbox{Im} \mathcal{I}_{u} (j) \mbox{Im} \mathcal{F}_{u}\left(q_{L}, p^\prime\right)}{E_{u} - E_{0}} 
+ \sum_{j^{\prime}=1}^{M-1} \chi(j^{\prime}) \sum_{u} 
\frac{\mbox{Im} \mathcal{I}_{u}(j) \mbox{Im} \mathcal{I}_{u} \left(j^{\prime}\right) }{E_{u} - E_{0}} 
 = 0 ,
\label{eq-Lagrange-J}  
\end{equation}  
Once they are determined, the relevant expansion coefficients
can be computed as
\begin{equation}  
A_{u} = \frac{2}{E_{u} - E_{0}}
\left[
-\frac{\mathcal{W}_u}{2} + 
\sum_{p} \lambda(p) \mathcal{F}_{u}\left(q_{L}, p\right)
+ \sum_{j=1}^{M-1} \chi(j) \mathcal{I}_{u}(j)
\right] ,
\label{eq-A_u}  
\end{equation}  
which enables to determine the electric current via Eq.\ \ (\ref{eq-J-lra}).
Notice that because $\mathcal{W}_{u}$ enters linearly Eqs.\ 
(\ref{eq-Lagrange-FW}), (\ref{eq-Lagrange-J}), and (\ref{eq-A_u}), the current computed 
via Eq.\ (\ref{eq-J-lra}) is proportional to the applied bias $V$.
That is, the minimization procedure described above indeed 
yields a solution corresponding to the linear response limit.
\section{Model}
\label{sec:model}
The method exposed in Sec.\ \ref{sec:method} is general. 
It will be applied to a concrete system.
The physical system considered in this paper consists of a single quantum 
dot connected to two noninteracting leads. It can be described by the Anderson impurity model 
\begin{eqnarray}  
H & = & 
\varepsilon_{L} \sum_{l=-1}^{-M_L}\sum_{\sigma}  a_{l,\sigma}^{\dagger} a_{l,\sigma}^{}
- t_{L} \sum_{l=-1}^{-M_L+1}\sum_{\sigma} \left(
  a_{l,\sigma}^{\dagger} a_{l-1,\sigma}^{} 
+ a_{l-1,\sigma}^{\dagger} a_{l,\sigma}^{}
\right) \nonumber \\ 
& &
+ \varepsilon_{R} \sum_{l=1}^{M_R}\sum_{\sigma}  a_{l,\sigma}^{\dagger} a_{l,\sigma}^{}
- t_{R} \sum_{l=1}^{M_R-1}\sum_{\sigma} \left(
  a_{l,\sigma}^{\dagger} a_{l+1,\sigma}^{} 
+ a_{l+1,\sigma}^{\dagger} a_{l,\sigma}^{}\right) \nonumber \\[-0.9ex]
& & \label{eq-hamiltonian}  \\[-0.9ex]
& & 
- t_{d,L} \sum_{\sigma} \left(a_{-1,\sigma}^{\dagger} d_{\sigma}^{} + d_{\sigma}^{\dagger} a_{-1,\sigma}^{}\right)
- t_{d,R} \sum_{\sigma} \left(a_{+1,\sigma}^{\dagger} d_{\sigma}^{} + d_{\sigma}^{\dagger} a_{+1,\sigma}^{}\right) 
\nonumber \\
& &
+ \varepsilon_{g}  \sum_{\sigma} d_{\sigma}^{\dagger} d_{\sigma}^{}
+  U \hat{n}_{d,\uparrow}^{} \hat{n}_{d,\downarrow}^{}, \nonumber
\end{eqnarray}  
where $a_{l,\sigma}$ ($a^{\dagger}_{l,\sigma}$) denote creation   
(annihilation) operators for electrons of spin $\sigma$ in the leads,  
 $d_{\sigma} \equiv a_{0,\sigma}$ ($d^{\dagger}_{\sigma}  \equiv a^{\dagger}_{0,\sigma}$) 
creates (destroys) electrons in the QD, and 
$\hat{n}_{d,\sigma} \equiv d_{\sigma}^{\dagger} d_{\sigma}^{}$
are electron occupancies per spin direction.
We shall consider $M_L = M_R \equiv M$, 
$t_{L} = t_{R} \equiv t$, $\varepsilon_{L} = \varepsilon_{R}$ 
(chosen as zero energy hereafter), 
$t_{d,L} = t_{d,R} \equiv t_{d}$, and this ensures the spatial inversion 
assumed above. The dot energy $\varepsilon_g$ can be tuned by varying 
a gate potential.
In view of the particle-hole symmetry of model (\ref{eq-hamiltonian}), 
($\varepsilon_g = - U/2$ is the particle-hole symmetric point), 
we can restrict ourselves to the range $\varepsilon_g > - U/2$. 
The number of electrons $N$ will be assumed 
equal to the number of sites ($N=2M+1$).
The electric potential entering Eq.\ (\ref{eq-W}) will be assumed 
constant within the reservoirs $\varphi_{l} = - \varphi_{-l} = - V/2$ for 
$0 < l < M$ and zero at the dot location, $\varphi_{0} = 0$.
\par
In isolated electrodes ($t_{d}=0$), the single-particle energies lie 
symmetrically around $\varepsilon = 0$.
Therefore, in order to eliminate energy corrections $\sim t/M$ in small clusters, 
it is advantageous to consider reservoirs with an 
odd number of sites $M$.\cite{Chiappe:03,Al-hassanieh:06} In the noninteracting case ($U=0$), 
the resonant tunneling at the Fermi energy ($\mu = 0$) is favored for odd $M$.
In the presence of interaction, the Kondo effect occurs when the dot spin couples with electrons 
of the leads at the Fermi level. 
This coupling is favored if the leads possess single electron levels of zero energy, 
and this is the case for reservoirs consisting 
of an odd number ($M$) of sites.
\par
Although the numerical values used for the results presented in 
this or in the next Sec.\ \ref{sec:uncorrelated} do not 
represent a special choice, we prefer to employ values   
corresponding to a possible physical realization of 
the above model, namely chains of QDs of silver. Such QDs were experimentally 
fabricated.\cite{Collier:97,Heath:97a,Markovich:98,Shiang:98,Medeiros:99,Henrichs:00,Sampaio:01,Beverly:02}
Their properties can be tuned in wide ranges by varying the dot diameter ($2R$) and interdot spacing 
($D$),\cite{Collier:97,Heath:97a,Markovich:98,Shiang:98,Medeiros:99,Henrichs:00,Sampaio:01,Beverly:02}, 
and the parameters are well documented in a series of experimental and theoretical 
works.\cite{Collier:97,Remacle:98b,Medeiros:99,Baldea:2002} For interdot spacings 
up to say, $D/2R\alt 1.3$, electron correlations are not important 
\cite{Baldea:2002,Baldea:2007,Baldea:2008}. Therefore, the reservoirs can be modeled, \emph{e.\ g.}, 
by chains of nearly touching QDs ($D/2R \simeq 1.07$, $t=1$\,eV). 
\par
In principle, the boundaries could be chosen at $ 0 < q_{R} = -q_{L} < M$.
Unless otherwise specified, we shall choose $q_{L,R}=\mp 2$.
It amounts to consider the central part to consist of the QD and two additional 
sites, one at either side of the ``device'', represented by the QD of interest.  
This bears the most resemblance to usual quantum chemical approaches to molecular transport, 
wherein the smallest possible parts of electrodes are included in the central part.
In Eq.\ (\ref{eq-wigner}), there are $n_{q_{L}} = 2 (M + q_{L}) + 1$ values of $r$. 
such that the values of site indices are $q_{L} \pm r = 0, \pm 1, \ldots, \pm (M + q_{L})$. 
According to the careful analysis of Ref.\ \onlinecite{Frensley:91}, the values of $p$ 
in Eq.\ (\ref{eq-wigner}) should span the first Brillouin zone of the reciprocal space 
of $2r$. 
\section{Conductance through a point contact in the absence of correlations}
\label{sec:uncorrelated}
Let us start with the noninteracting case, amounting to switch off the 
Coulomb interaction ($U=0$) in Eq.\ (\ref{eq-hamiltonian}). 
This represents the textbook case of conduction through a 
single level system.\cite{Datta:05} 
By approaching the resonance, 
$\varepsilon_{g} \to 0$, the transmission becomes perfect.
The curve of the conductance $G(\varepsilon_{g})$ exhibits a 
peak characterized by a height $G(0) = G_{0}$ and a half-width 
parameter $\Gamma = 2 t_{d}^2/t$.
\par
For the numerical calculations in this section, we shall choose a value 
$t_{d}/t=0.4$. This corresponds to an Ag-QD chain, 
with a QD in the middle slightly more distant 
($D^{\prime}/2R = 1.24$) from its two neighbors than the other QDs in the chains 
($D/2R = 1.07$, \emph{cf.\ }Sec.\ \ref{sec:model}).
\par
In the case of identical reservoirs and contacts, the zero-bias 
conductance can be obtained from the Friedel-Langreth sum rule 
\cite{Langreth:66,Shiba:75,Meir:92,Meir:94}
\begin{equation}  
G/G_0 = \sin^2(\pi n_{d}/2),
\label{eq-friedel}  
\end{equation}  
where $G_{0} \equiv 2 e^2/h$ is the conductance quantum. 
The dot occupancy per spin direction 
$n_{d} = \sum_{\sigma} \langle \Psi_{0}\vert \hat{n}_{d,\sigma}\vert \Psi_{0}\rangle$ 
will be computed by numerical exact diagonalization.
\par
\begin{figure}[htb]
\centerline{\hspace*{-0ex}
\includegraphics[width=0.42\textwidth,angle=-90]
{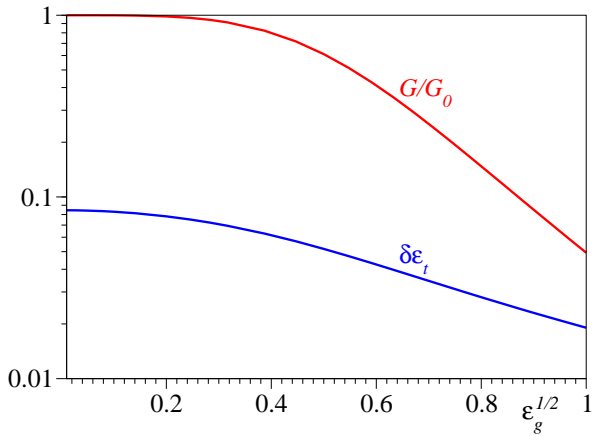}
}
\caption{(Color online) Results for the normalized conductance $G/G_0$ 
and lowest excitation energy $\delta \varepsilon_t$ (tunneling splitting)
obtained by exact diagonalization in the uncorrelated system ($U=0$) for chains 
with 63 sites, $t=1$, and $t_{d}=0.4$. Note the 
logarithmic scale on the ordinate and the square root of the dot energy(=energy barrier) 
$\varepsilon_g$ on the abscisa. The linearity of the two curves at larger $\varepsilon_g$ confirms 
the interpretation within the tunneling model.}
\label{fig_tunnel_interpretation_landauer}
\end{figure}

In Fig.\ \ref{fig_tunnel_interpretation_landauer}, we present numerical results 
obtained for 63 sites by means of Eq.\ (\ref{eq-friedel}), which 
show a peak in the conductance $G(\varepsilon_{g})$ with a half width at half maximum 
in very good agreement with the formula $\Gamma = 2 t_{d}^2/t$.
These results are in accord with the fact that 
the single-particle quantum tunneling constitutes the underlying physical phenomenon: 
At sufficiently large 
values of $\varepsilon_{g}$, the curve for $\log G(\varepsilon_{g})$ in Fig.\ 
\ref{fig_tunnel_interpretation_landauer} varies linearly with $\varepsilon_{g}^{1/2}$,
as expected for the transmission coefficient through an energy barrier 
$\varepsilon_{g}$. The lowest excitation energy, also shown 
in Fig.\ \ref{fig_tunnel_interpretation_landauer}, displays a similar dependence,
which confirms that it plays the role of a tunneling splitting energy.  
\par
The curve of conductance for chains with 63 sites depicted in 
Fig.\ \ref{fig_tunnel_interpretation_landauer} is very close to the exact result 
for infinite chains presented in Fig.\ 5 of Ref.\ \onlinecite{Al-hassanieh:06}.
It has been shown there that the latter result agrees very well 
with the time-dependent-DMRG result for 64 sites.\cite{Al-hassanieh:06} 
The size $N=63$ chosen by us is the closest to chain size $N=64$ 
used in the time-dependent DMRG-calculations compatible with 
with our choice ($N=2M+1$, with odd $M$, \emph{cf.\ }Sec.\ \ref{sec:model}). 
\par
We shall now present the results of the SWF-method discussed in Sec.\ \ref{sec:method}.
From the point of view of the computation time, large chain sizes are more prohibitive 
for the SWF-method than for exact diagonalization. 
Indeed, by inspecting Eqs.\ 
(\ref{eq-J-lra}), (\ref{eq-Lagrange-FW}), (\ref{eq-Lagrange-J}), and (\ref{eq-A_u}) 
one can see that the computing time for $J$ scales as $N^6$ in the noninteracting case, 
where the number of relevant excitations scales as $N^2$.
Because the Hamiltonian (\ref{eq-hamiltonian}) is quadratic in the noninteracting case, 
exact numerical diagonalization can be straightforwardly carried out even for very 
long chains, \emph{e.\ g.}, much longer than those that can be handled 
by the time-dependent DMRG.
\par
\begin{figure}[htb]
\centerline{\hspace*{-0ex}
\includegraphics[width=0.42\textwidth,angle=-90]
{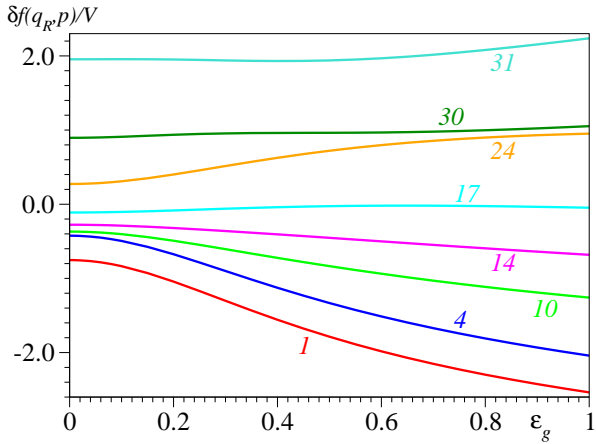}
}
\caption{(Color online) The difference  
$\delta f(q_{R}, p) \equiv  f(q_{R}, p) -  f_{0}(q_{R}, p)$ 
(in arbitrary units) plotted versus gate potential $\varepsilon_{g}$
for 63 sites, $t=1$, $t_{d} = 0.4$. 
The nonvanishing values at the right interface $q_{R}$ and positive 
momenta $p$  
indicate that the Wigner distribution function of outgoing electrons  
is free. The values of $k$ ($p = k \pi/n_{q_{L}}$) are given in the legend.}
\label{fig_dFW}
\end{figure}

Fig.\ \ref{fig_dFW} represents a counterpart of the bottom panel of Fig.\ 1 of Ref.\ 
\onlinecite{DelaneyGreer:04a} and illustrates a characteristic feature of the SWF-method discussed in 
Sec.\ \ref{sec:method}: the reservoirs only constrain the Wigner distribution 
function of incoming electrons, while the distribution 
of outgoing electrons is free. For instance, at the right interface 
$\delta f(q_{R}, p) \equiv  f(q_{R}, p) -  f_{0}(q_{R}, p)$ 
is zero for $p<0$ (in accord with Eq.\ (\ref{eq-FW-R})) 
but has nonvanishing values for $p > 0$. 
Curves for the latter case are shown in Fig.\ \ref{fig_dFW}, 
depicted for several positive values of $p$. 
\par
\begin{figure}[htb]
\centerline{\hspace*{-0ex}
\includegraphics[width=0.42\textwidth,angle=-90]
{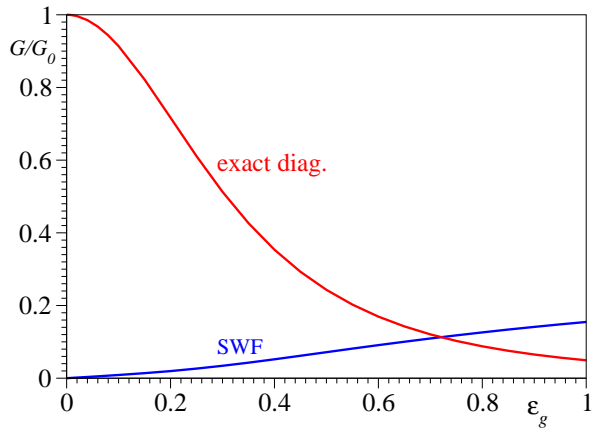}
}
\caption{(Color online) Results on the normalized conductance $G/G_0$ 
obtained by exact diagonalization and the method based on the Wigner function 
for 63 sites in the absence of correlations, $U=0$ (same parameters as in
Fig.\ \ref{fig_tunnel_interpretation_landauer}).}
\label{fig_sigma_landauer_vs_dg}
\end{figure}

We shall now compare the exact results with those of the SWF-method. 
In Fig.\ \ref{fig_sigma_landauer_vs_dg}, the SFW-curve for zero-bias conductance 
is plotted along with the exact curve. As one can clearly see there, the SFW-curve
looks completely different, bearing no resemblance with the exact curve.
Most unphysically, the SWF-conductance vanishes for resonant tunneling 
($\varepsilon_{g}=0$), where it should attain the maximum value $G=G_{0}$.
\par
\begin{figure}[htb]
\centerline{\hspace*{-0ex}
\includegraphics[width=0.42\textwidth,angle=-90]
{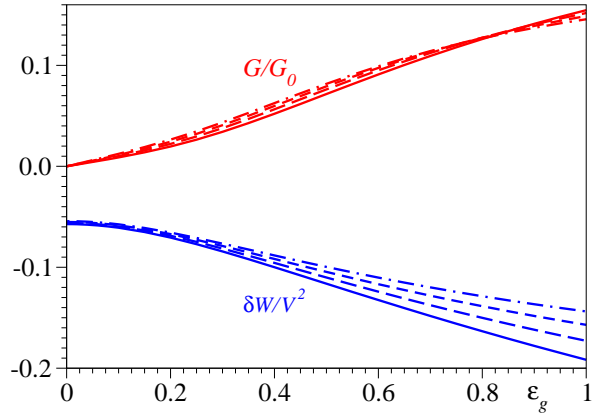}
}
\caption{(Color online) Results for the conductance $G/G_0$ 
and the bias-induced condensation energy $\delta \mathcal{W}$ 
(the latter in arbitrary units) 
obtained by means of the SWF-method for boundaries chosen at 
$q_{R} = - q_{L} =2, 3, 4, 5$ (values increasing upwards for the $G$-curves and downwards for 
the $\delta\mathcal{W}$-curves at, say, $\varepsilon_{g} = 0.5$). 
Noteworthy, the conductance 
does not sensitively depend on the choice of boundaries.}
\label{fig_various_Mp_qL}
\end{figure}

To compute the SWF-curve of Fig.\ \ref{fig_sigma_landauer_vs_dg}, we have chosen 
$q_{L}=-2$ and $q_{R}=2$.  
A nontrivial realistic \emph{ab initio} calculation is so demanding, that, 
besides the device (a molecule or a few QDs), 
if at all, only small parts of the electrodes 
can be accounted for in the heaviest part of the computation. 
Therefore, practically no or very limited freedom 
remains to choose the boundaries $q_{L,R}$.
The fact that in the present case the eigenvalue problem can be solved exactly for large systems 
$\mathcal{S}$ enables us to flexibly change $q_{L,R}$ and to inspect 
the impact on the solution, and, maybe to make it having some resemblance to the exact one.
In Fig.\ \ref{fig_various_Mp_qL}, we present results derived by choosing different $q_{L,R}$. 
For all choices, the optimization yields minimum values of the total energy $\mathcal{E}$, 
Eq.\ (\ref{eq-total-energy}), below the value $\mathcal{E}_{0}$ corresponding to the 
trivial situation 
$A_{n} = 0, \forall n \neq 0$, \emph{i.\ e.,} the ``condensation'' energy 
$\delta \mathcal{W} \equiv \mathcal{E} - \mathcal{E}_{0}$ is negative.
However, the conductance changes only by factors of the order of unity.
Definitely, it cannot be made more akin to the exact $G$. 
\par
\begin{figure}[htb]
\centerline{\hspace*{-0ex}
\includegraphics[width=0.42\textwidth,angle=-90]
{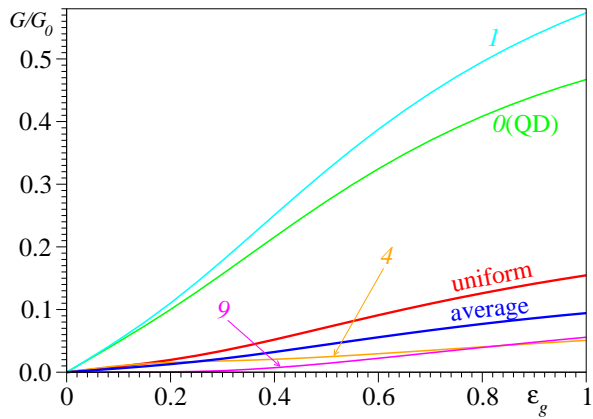}
}
\caption{(Color online) SFW-results for the conductance $G/G_0$ of 
the chain with 63 sites at $t=1$ and $t_{d}=0.4$. 
The curves computed without imposing the continuity equation 
exhibit a strong site-dependent current and substantially depart 
from that labelled by \emph{uniform}, computed by imposing this equation. 
Nor the average taken along the chain (label \emph{average}) 
represents a satisfactory approximation of the latter.
The number of site $q$ in the chain is specified in the legend.  
The dot is located at $q=0$.
}
\label{fig_factJ}
\end{figure}

The curves presented in the above Figs.\ \ref{fig_dFW}, \ref{fig_sigma_landauer_vs_dg}, and 
\ref{fig_various_Mp_qL} have been deduced by constraining the solution 
to satisfy the equation of continuity, as discussed in Sec.\ \ref{sec:method}.
In Fig.\ \ref{fig_factJ}, results obtained by 
imposing the equation of continuity (depicted by the line denoted by 
\emph{uniform}) are compared with those derived without 
imposing this equation. As visible there, the values of the electric current do 
display a significant dependence on site. 
The latter is indicated by the numbers in the legend of Fig.\ \ref{fig_factJ}.
Neither the value of the curent through  
the QD nor the average along the chain (label $0$ and \emph{average}, respectively)
coincides or reasonably approximates that deduced by imposing the equation of continuity.
It was claimed that, although in principle necessary because of certain approximations 
(see Sec.\ \ref{sec:discussion} for more details), there was no need to impose the equation of 
continuity for the calculations reported in Refs.\ \onlinecite{DelaneyGreer:04a,DelaneyGreer:04b}.
Obviously, the fact that in our case the minimization without imposing Eq.\ (\ref{eq-j-average})
leads to a solution for $\Psi$ that violates the equation of continuity is not 
the result of any approximation: 
except for the SWF-method itself, our results are affected by no further approximation.
To conclude, Fig.\ \ref{fig_factJ} demonstrates 
that the SWF-method does not automatically 
satisfy the continuity equation, not even approximately. 
\section{Conductance through a point contact in the presence of correlations}
\label{sec:correlated}
\begin{figure}[htb]
\centerline{\hspace*{-0ex}
\includegraphics[width=0.42\textwidth,angle=-90]
{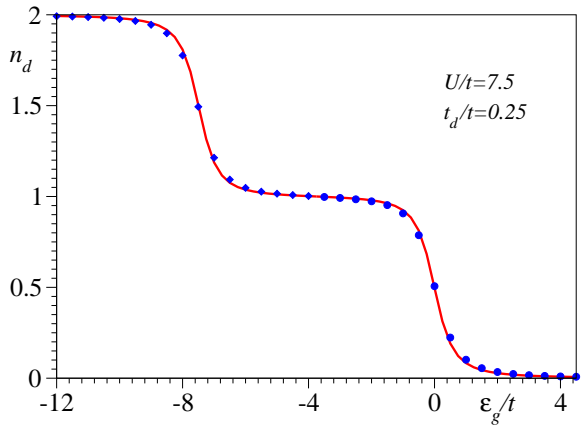}
}
\caption{(Color online) Dot occupancy $n_{d}$ computed by exact diagonalization 
for 7- and 11-site clusters (solid lines and points, respectively) 
plotted versus dot energy $\varepsilon_g$. Notice the small difference between 
the results for 7 and 11 sites, indicating that finite-size effects play a reduced role.}
\label{fig_sigma_n_dot_7_and_11_sites_plateaus}
\end{figure}

In this section we shall apply the SWF-method exposed in Sec.\ \ref{sec:method}
for the case of nonvanishing $U$, where correlations are known to play an important role.
The physics of model (\ref{eq-hamiltonian}) with $U\neq 0$ is also well understood.\cite{Ng:88}
For strong interaction ($U$) and low temperatures, by varying the gate potential $\varepsilon_{g}$, 
one observes plateaus of well defined dot charge, corresponding 
to a dot that is empty, singly, and doubly 
occupied: $n_{d}=0$ ($0 < \varepsilon_{d}$), $n_{d}=1$ ($ -U < \varepsilon_{d} < 0$),
and $n_{d}=2$ ($\varepsilon_{d} < -U$), respectively.
This behavior can be demonstrated by exact numerical diagonalization in 
small clusters, as illustrated by the curves of Fig.\ \ref{fig_sigma_n_dot_7_and_11_sites_plateaus}.
\par
For low temperatures but above the so-called Kondo temperature $T_K$ 
(often much smaller than 1\,K), 
the conductance $G(\varepsilon_g)$ exhibits two narrow Coulomb blockade peaks 
located at $\varepsilon_g=-U$ and $\varepsilon_g=0$, while in between 
it almostly vanishes (``Coulomb valley''). By decreasing the temperature below $T_K$,  
correlation effects in the singly occupied state yields a sharp (Kondo) resonance 
in the density of states at the Fermi level, and this 
gives rise to a characteristic plateau of width $\sim U$ 
in the curve of $G$ versus $\varepsilon_g$.
In the middle of the Coulomb valley ($\varepsilon_g = - U/2$), perfect transmission 
occurs, leading to the ideal conductance value $G_0=2e^2/h$ (unitary limit).
\par
Numerous results obtained for the model (\ref{eq-hamiltonian}) by considering semi-infinite 
leads were published in the literature; see, \emph{e.\ g.}, Refs.\ 
\onlinecite{Izumida:01,Costi:01,Al-hassanieh:06}. 
A comparison between the results for semi-infinite leads 
and the SWF-method would make no sense. 
At zero temperature, the case for which the SWF-method was developed, 
in the range $-U \alt \varepsilon_{g} \alt 0$ for the realistic case of 
semi-infinite electrodes the conductance is dominated 
by the Kondo plateau. Its formation requires chain  
sizes larger than the Kondo cloud, which extends over a number of sites 
$\xi_{K} \sim t/T_{K}$. The latter rapidly grows (exponentially for large $U$)
beyond the sizes, which neither the exact diagonalization, nor the SWF, or often even the  
DMRG approach \cite{Al-hassanieh:06} can handle.
No Kondo peak can be formed for chains shorter than the Kondo screening length $\xi_{K}$.
However, in short chains the $G(\varepsilon_g)$-curve should still display the 
Coulomb blockade peaks at $\varepsilon_g=0$ and $\varepsilon_g=-U$.
\par
Although the clusters considered in this section comprise a small number of sites, 
the results are significant. Since our main purpose here is to address 
the issue of the validity of the SWF-method, we do not intend to discuss 
finite-size effects here. However, we do not expect that they are essential:
the inspection of Fig.\ \ref{fig_sigma_n_dot_7_and_11_sites_plateaus} 
reveals that the differences between chains with seven 
($M=3$) and eleven ($M=5$) sites are not substantial.
\par
\begin{figure}[htb]
\centerline{\hspace*{-0ex}
\includegraphics[width=0.42\textwidth,angle=-90]
{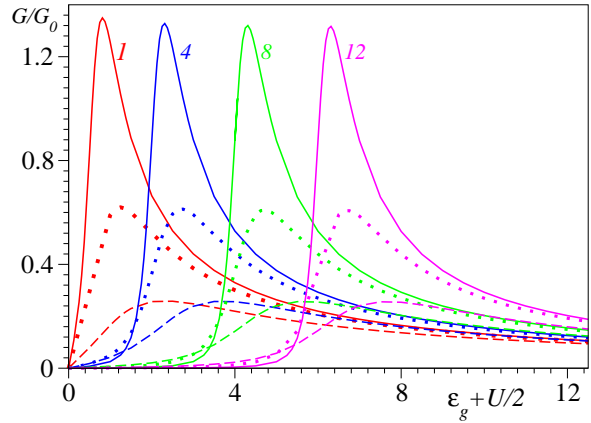}
}
\caption{(Color online) Coulomb blockade peaks of the zero-bias conductance 
for 7-site clusters as predicted by the SWF-method. 
The solid, dotted, and dashed lines correspond to the values $t_{d}=0.125$, $0.25$, and $0.5$ 
respectively. The values of $U$ are given in the legend, and $t=1$. Notice that, unphysically, 
the maximum conductance is predicted to increase with decreasing $t_{d}$ 
even beyond the ideal value $G_{0}$ ($G/G_{0} > 1$).}
\label{fig_Coulomb_blockade}
\end{figure}

The primary reason why we restrict ourselves to chains with seven sites is 
technical, but this also rises supplementary doubts on the applicability of the SWF-methods 
for systems of interest for molecular electronics.
Eq.\ (\ref{eq-Psi}) is an expansion over the 
complete set of eigenstates of $H$ and, according to Eq.\ (\ref{eq-J-lra}), 
in principle all eigenstates of odd parity contribute to the current. 
Calculations contradict the naive expectation that in the linear response 
approximation only a reduced number of excited states are important. 
For the couplings $U$ employed in Fig.\ \ref{fig_Coulomb_blockade}, out of a total of 1225 states 
with spin projection $S_z=+1/2$, the first 300 states are not always enough to reach convergence.
For eleven-site chains (the next larger size of interest), there are 213444 eigenstates with $S_z=+1/2$.
One would probably need to target many thousands thereof in order to get the matrix elements 
necessary for convergent results.
For this formidable task, one should run the Lanczos procedure 
three times, and this separately for each of the matrix elements entering Eqs.\ 
(\ref{eq-F}), (\ref{eq-J}), and (\ref{eq-P}),
in a manner similar to but more involved than that employed to compute frequencies and intensities 
of optical lines.\cite{koeppel:84,Baldea:97,Baldea:2007,Baldea:2008}   
The method of only computing convoluted spectra by means of the continued fraction 
algorithm \cite{HHK:80,fulde:91,dagotto:94} is inadequate for this purpose.
\par
Our results on the conductance computed by means of the SWF-method are 
collected in Fig.\ \ref{fig_Coulomb_blockade}. 
(Notice that only the halves of the curves situated at the right of the 
particle-hole symmetric point $\varepsilon_{g} + U/2 =0$ are shown.)
As one can see there, the 
curves for the conductance, calculated for several values of $U$ and $t_{d}$, 
exhibit maxima at the gate potential values 
$\varepsilon_{g} \agt 0$
($\varepsilon_{g} + U/2 \agt U/2$), 
\emph{i.\ e.,} at the position where the Coulomb blockade peaks are expected. 
Their distance from the ideal Coulomb blockade location increases with 
$t_{d}$, a behavior similar to that of their width. While this behavior is physically plausible, 
unfortunately, 
the prediction of the SWF-method for the height of the peaks is quite unphysical:
the height is found to vary roughly inversely proportional to $t_{d}$. 
The consequence of this dependence
is that, as visible in Fig.\ \ref{fig_Coulomb_blockade}, the height of the $G$-peak even 
attains values exceeding the ideal value $G_{0}$. To conclude this section,  
the results of the SWF-method are unphysical; the conductance of 
correlated systems evaluated by this method cannot be trusted.   
\section{Discussion}
\label{sec:discussion}
We have two comments on the SFW-method. 
They concern this method in general, and are not related to the linear response limit.
Firstly, in Ref.\ \onlinecite{DelaneyGreer:04b}, it was claimed that the imposition of 
Eqs.\ (\ref{eq-j-average}) is necessary, because the continuity equation 
$\partial \rho/\partial t + \mathbf{\nabla} \raisebox{0.7ex}{.}  \mathbf{j} = 0 $, 
derived from the Schr\"odinger equation in the presence of local interactions, 
does not hold for nonlocal interactions. Accordingly, 
position-dependent currents could be an effect of \emph{ab initio} 
molecular electronics calculations employing nonlocal effective core 
potentials or pseudopotentials, or a result of truncating 
the molecular orbital basis set or the CI (configuration interaction).
While all these may in general be sources of violating the continuity equation,
for the SWF-method there still exists another reason to impose the 
constrains (\ref{eq-j-average}) in a steady state. 
The wave function $\Psi$ determined by means of the SWF-method 
does not represent an eigenstate of the Hamiltonian. $\Psi$ is time independent 
only because the formalism is time independent, and not the result 
\emph{e.\ g.}, of taking the limit $t\to \infty$ to get a steady state.
No demonstration has been given in the works dealing with the SWF-method 
\cite{DelaneyGreer:04a,DelaneyGreer:04b,DelaneyGreer:06,Fagas:06,Fagas:07}
that the minimization procedure without imposing Eqs.\ (\ref{eq-j-average}) yields a 
solution compatible with the continuity equation. 
While the imposition of  Eqs.\ (\ref{eq-j-average}) is seemingly unnecessary for 
the case considered in Ref.\ \onlinecite{DelaneyGreer:04b}, there is no rationale 
for this in general. For illustration, we have presented a counter-example in 
Sec.\ \ref{sec:uncorrelated}. The equation of continuity 
$ \partial n_{l} / \partial t = (i/\hbar)[n_{l}, H] = - j_{l} + j_{l-1}$ 
(\emph{cf.\ }Ref.\ \onlinecite{Caroli:71}) holds for the model Hamiltonian 
(\ref{eq-hamiltonian}), and nevertheless the current 
computed without imposing Eqs.\ (\ref{eq-j-average}) is strongly site-dependent.
\par
The second and more important point concerns the manner of imposing 
boundary conditions in Ref.\ \onlinecite{DelaneyGreer:04a}.
In the approaches based on the Wigner function 
within the single-particle approximation the influence of the applied electric field is 
accounted for both at the boundary conditions, via the shift $\mu_L - \mu_R = e V$
(\emph{cf.\ }Eqs.\ (\ref{eq-FW-LB}) and (\ref{eq-FW-RB})), and on the electron dynamics 
within the device.\cite{Frensley:91,Frensley:86,Frensley:88}
Within the methods based on the NEGF the applied voltage is usually considered solely via the 
the chemical potential imbalance, and many-body \cite{HaugJauho}
methods are employed to treat correlation effects due to 
interactions within the device without applied field. 
In both cases, the applied bias represents the driving force of current flow. 
In the SWF-method, the effect of the applied 
field at the boundaries is entirely neglected 
(\emph{cf.\ }Eqs.\ (\ref{eq-FW-L}) and (\ref{eq-FW-R})).
Let us suppose that (i) we would be able to reliably solve the minimization problem 
as prescribed by the SWF-method and exactly (or at least very accurately) 
determine $\Psi$ for very large 
systems, including large parts of the reservoirs (assumed identical), 
and (ii) in the latter the single-particle description applies.
Then, according to Eqs.\  (\ref{eq-FW-L}) and (\ref{eq-FW-R}), 
the Wigner functions at the boundaries would reduce to the 
Fermi distribution functions of the left and right reservoirs 
characterized by the \emph{same} chemical potential $\mu$. 
This would imply that there would be no difference between incoming 
electrons from the left and right reservoirs.
Then, it is not at all surprising \emph{e.\ g.\ }that 
in the extreme case, where all sites are noninteracting and identical
($U=0$ and $t_{d} = t$), instead of being maximum, $G=G_{0}$, 
the conductance vanishes for $\varepsilon_{g}=0$, 
as visible for the SWF-curves of Figs.\ \ref{fig_sigma_landauer_vs_dg},
\ref{fig_various_Mp_qL}, and \ref{fig_factJ}. 
In reality, the correct wave function $\Psi$ describing the 
steady state current flow should yield a Wigner function that reduces at the boundaries 
to the Fermi distribution functions characterized by different chemical potentials 
$\mu_{L,R} = \mu \pm eV/2$. 
\par
Since the above analysis reveals that the boundary conditions 
(\ref{eq-FW-L}) and (\ref{eq-FW-R}) are inadequate, attempting to mend 
the SWF-method would be desirable. In view of the above considerations, perhaps 
the most natural attempt would
be to modify Eqs.\ (\ref{eq-FW-L}) and (\ref{eq-FW-R}) by using 
instead of $\Psi_{0}$ the ground state 
$\Phi_{0}$ of the system in the presence of the applied potential, \emph{i.\ e.},
$H_{T}\vert\Phi_{0}\rangle = E_{T,0} \vert\Phi_{0}\rangle$. 
Although the Wigner functions entering the r.h.s.\ of Eqs.\ (\ref{eq-FW-L}) and (\ref{eq-FW-R}), 
calculated by using $\Phi_{0}$ istead of $\Psi_{0}$, do not necesarily reduce to the left and right 
Fermi distributions, this procedure would at least account 
for the chemical potential imbalance at the boundaries. 
However, as revealed by a straightforward analysis, this modification does not 
yield the desired improvement: the solution of the minimization is just $\Psi = \Phi_{0}$. 
This solution obviously satisfies the boundary constraints (\ref{eq-FW-L}) and (\ref{eq-FW-R})
as well as the continuity equation 
(electrical perturbations only depend on electron density), 
and, being the ground state of $H_{T}$, it trivially minimizes $\mathcal{E}$ of 
Eq.\ \ref{eq-total-energy}. This result is obviously general, 
\emph{i.\ e.}, it holds beyond the linear response approximation.
Still, as a verification, we have performed the modification 
$\Psi_{0} \to \Phi_{0}$ in the r.h.s.\ of Eqs.\ (\ref{eq-FW-L}) and (\ref{eq-FW-R}),
and carried out straightforward calculations within the linear response approximation.
They yield $\vert \Psi \rangle = 
\vert \Psi_{0}\rangle + \sum_{n\neq 0} \mathcal{W}_{n}/(E_{0} - E_{n}) \vert \Psi_{n}\rangle $, 
and in the r.h.s.\ one immediately recognizes the ground state $\Phi_{0}$ 
of $H_{T} = H + W$ in the first-order of perturbation theory.
Unless the system is superconducting, 
the current (conductance) vanishes in the state described by 
the wave function $\Psi = \Phi_{0}$. So, even with this ``remedy'', the SWF-method is unable 
to describe electric transport through nanosystems.
\section{Conclusion}
\label{sec:conclusion}
Several recent studies proposed and applied a many-body time-independent 
method to compute steady-state electric transport 
in molecular systems, whose key ingredient was the formulation of the 
boundary conditions in terms of the Wigner distribution 
function.\cite{DelaneyGreer:04a,DelaneyGreer:04b,DelaneyGreer:06,Fagas:06,Fagas:07}
The fact that this approach yielded values of the electric current through 
molecules comparable with 
those measured in experiments, which are usually orders of magnitudes lower than 
the predictions of other theoretical treatments, 
was considered very encouraging.
However, the mere fact that a theoretical method compares favorably with 
experiments cannot be taken as support for its correctness. It should also be able 
to correctly reproduce well established results. 
\par
In this paper, we have presented results demonstrating that the SWF-method 
is unable to reliably evaluate the zero-bias conductance of the 
simplest uncorrelated and correlated systems of interest for molecular 
and nanoscopic systems, namely a single QD. It fails to retrieve 
the result $G=G_{0}$ for resonant tunneling through a single QD without correlations, 
where it predicts a vanishing conductance instead. In the presence of correlations, 
the conductance at the peaks of Coulomb blockade is unphysically predicted to increase 
with decreasing dot-electrode coupling ($t_{d}$) and can even exceed the conductance
quantum $G_{0}$.
\par
While the idea of formulating 
boundary conditions in terms of the Wigner function for correlated many-body systems
is interesting, the manner in which it was imposed in Ref.\ \onlinecite{DelaneyGreer:04a} 
turns out to be inappropriate. 
It misses the fact that, in accord with our physical understanding, 
the current flow is due to an \emph{asymmetric} injection of  
electrons from reservoirs into the device, 
and that injected electrons are very well described by Fermi distributions 
with different chemical potentials. 
Moreover, as results from the analysis at the end of Sec.\ \ref{sec:discussion}, 
unfortunately there is no simple remedy of the SWF-method; the modification of  
the boundary conditions \emph{in the spirit} of Ref.\ \onlinecite{DelaneyGreer:04a} 
such as to account for a nonvanishing chemical potential shift does not yield the desired improvement. 
\par
In addition, as a side note, 
we believe that the very fact that astonishingly numerous 
states with high excitation energies are found to contribute to the conductance 
in the \emph{linear approximation} is an indication that the SWF-method, 
even if it were physically sound, would be of little pragmatical use 
for strongly correlated nanosystems. Because it would be hardly conceivable that 
\emph{ab initio} calculations for real molecular systems,
by far more complex than the presently considered model,
could provide a wave function $\Psi$ with the accuracy needed for 
reaching reliable convergent results. 
These considerations raise doubts on the possibility to develop a 
viable SWF-approach for the electric transport through nanoscopic and 
molecular systems. 
\section*{Acknowledgments}
The authors acknowledge with thanks the financial support for this work 
provided by the Deu\-tsche For\-schungs\-ge\-mein\-schaft (DFG).


\begin{thebibliography}{10}

\bibitem{Wenzel:02}
J.~Heurich, J.~C. Cuevas, W.~Wenzel, and G.~Sch\"on,
\newblock Phys. Rev. Lett. {\bf 88}, 256803 (2002).

\bibitem{Nitzan:03}
A.~Nitzan and M.~A. Ratner,
\newblock Science {\bf 300}, 1384 (2003).

\bibitem{Stokbro:03}
K.~Stokbro, J.~Taylor, M.~Brandbyge, J.-L. Mozos, and P.~Ordejón,
\newblock Comp. Mat. Sci. {\bf 27}, 151 (2003).

\bibitem{Reed:97}
M.~A. Reed, C.~Zhou, C.~J. Muller, T.~P. Burgin, and J.~M. Tour,
\newblock Science {\bf 278}, 252 (1997).

\bibitem{Xiao:04}
X.~Xiao, B.~Xu, and N.~Tao,
\newblock Nano Letters {\bf 4}, 267 (2004).

\bibitem{Daosh:05}
T.~Dadosh et~al.,
\newblock Nature {\bf 436}, 677  (2005).

\bibitem{tsutsui:06}
M.~Tsutsui, Y.~Teramae, S.~Kurokawa, and A.~Sakai,
\newblock Applied Physics Letters {\bf 89}, 163111 (2006).

\bibitem{Landauer:70}
R.~Landauer,
\newblock Phil. Mag. {\bf 21}, 863  (1970).

\bibitem{Buttiker:85}
M.~B\"uttiker, Y.~Imry, R.~Landauer, and S.~Pinhas,
\newblock Phys. Rev. B {\bf 31}, 6207 (1985).

\bibitem{HaugJauho}
H.~Haug and A.-P. Jauho,
\newblock {\em Quantum Kinetics in Transport and Optics of Semiconductors},
  volume 123,
\newblock Springer Series in Solid-State Sciences, Berlin, Heidelberg, New
  York, 1996.

\bibitem{Datta:05}
S.~Datta,
\newblock {\em Quantum Transport: Atom to Transistor},
\newblock Cambridge Univ. Press, 2005.

\bibitem{White:04}
S.~R. White and A.~E. Feiguin,
\newblock Phys. Rev. Lett. {\bf 93}, 076401 (2004).

\bibitem{Daley:04}
A.~J. Daley, C.~Kollath, U.~Schollw\"{o}ck, and G.~Vidal,
\newblock Journal of Statistical Mechanics: Theory and Experiment {\bf 2004},
  P04005 (2004).

\bibitem{White:05}
A.~E. Feiguin and S.~R. White,
\newblock Phys. Rev. B {\bf 72}, 020404 (2005).

\bibitem{Bulla:08}
R.~Bulla, T.~A. Costi, and T.~Pruschke,
\newblock Rev. Mod. Phys. {\bf 80}, 395 (2008).

\bibitem{Taylor:01}
J.~Taylor, H.~Guo, and J.~Wang,
\newblock Phys. Rev. B {\bf 63}, 245407 (2001).

\bibitem{Brandbyge:02}
M.~Brandbyge, J.-L. Mozos, P.~Ordej\'on, J.~Taylor, and K.~Stokbro,
\newblock Phys. Rev. B {\bf 65}, 165401 (2002).

\bibitem{Rocha:06}
A.~R. Rocha et~al.,
\newblock Phys. Rev. B {\bf 73}, 085414 (2006).

\bibitem{Wenzel:03}
M.~H. Hettler, W.~Wenzel, M.~R. Wegewijs, and H.~Schoeller,
\newblock Phys. Rev. Lett. {\bf 90}, 076805 (2003).

\bibitem{DelaneyGreer:04a}
P.~Delaney and J.~C. Greer,
\newblock Phys. Rev. Lett. {\bf 93}, 036805 (2004).

\bibitem{DelaneyGreer:04b}
P.~Delaney and J.~C. Greer,
\newblock Int. J. Quant. Chem. {\bf 100}, 1163 (2004).

\bibitem{Muralidharan:06}
B.~Muralidharan, A.~W. Ghosh, and S.~Datta,
\newblock Phys. Rev. B {\bf 73}, 155410 (2006).

\bibitem{Thygesen:07}
K.~S. Thygesen and A.~Rubio,
\newblock J. Chem. Phys. {\bf 126}, 091101 (2007).

\bibitem{Vignale:87}
G.~Vignale and M.~Rasolt,
\newblock Phys. Rev. Lett. {\bf 59}, 2360 (1987).

\bibitem{Vignale:88}
G.~Vignale and M.~Rasolt,
\newblock Phys. Rev. B {\bf 37}, 10685 (1988).

\bibitem{toher:07}
C.~Toher and S.~Sanvito,
\newblock Phys. Rev. Lett. {\bf 99}, 056801 (2007).

\bibitem{Ke:07}
S.-H. Ke, H.~U. Baranger, and W.~Yang,
\newblock J. Chem. Phys. {\bf 126}, 201102 (2007).

\bibitem{Fagas:07}
G.~Fagas and J.~C. Greer,
\newblock Nanotechnology {\bf 18}, 424010 (4pp) (2007).

\bibitem{Fagas:06}
G.~Fagas, P.~Delaney, and J.~C. Greer,
\newblock Phys. Rev. B {\bf 73}, 241314(R) (2006).

\bibitem{Frensley:91}
W.~R. Frensley,
\newblock Rev. Mod. Phys. {\bf 63}, 215 (1991).

\bibitem{Frensley:86}
W.~R. Frensley,
\newblock Phys. Rev. Lett. {\bf 57}, 2853 (1986).

\bibitem{Frensley:88}
W.~R. Frensley,
\newblock Phys. Rev. Lett. {\bf 60}, 1589 (1988).

\bibitem{Caroli:71}
C.~Caroli, R.~Combescot, P.~Nozi\`eres, and D.~Saint-James,
\newblock J. Phys. C: Solid State Physics {\bf 4}, 916 (1971).

\bibitem{Chiappe:03}
G.~Chiappe and J.~A. Verg\'{e}s,
\newblock J. Phys.: Condensed Matter {\bf 15}, 8805 (2003).

\bibitem{Al-hassanieh:06}
K.~A. Al-Hassanieh, A.~E. Feiguin, J.~A. Riera, C.~A. Busser, and E.~Dagotto,
\newblock Phys. Rev. B {\bf 73}, 195304 (2006).

\bibitem{Collier:97}
C.~P. Collier, R.~J. Saykally, J.~J. Shiang, S.~E. Henrichs, and J.~R. Heath,
\newblock Science {\bf 277}, 1978 (1997).

\bibitem{Heath:97a}
J.~Heath, C.~Knobler, and D.~Leff,
\newblock J. Phys. Chem. B {\bf 101}, 189 (1997).

\bibitem{Markovich:98}
G.~Markovich, C.~P. Collier, and J.~R. Heath,
\newblock Phys. Rev. Lett. {\bf 80}, 3807 (1998).

\bibitem{Shiang:98}
J.~Shiang, J.~Heath, C.~Collier, and R.~Saykally,
\newblock J. Phys. Chem. B {\bf 102}, 3425 (1998).

\bibitem{Medeiros:99}
G.~Medeiros-Ribeiro, D.~A.~A. Ohlberg, R.~S. Williams, and J.~R. Heath,
\newblock Phys. Rev. B {\bf 59}, 1633 (1999).

\bibitem{Henrichs:00}
S.~Henrichs, C.~Collier, R.~Saykally, Y.~Shen, and J.~Heath,
\newblock J. Amer. Chem. Soc. {\bf 122}, 4077 (2000).

\bibitem{Sampaio:01}
J.~Sampaio, K.~Beverly, and J.~Heath,
\newblock J. Phys. Chem. B {\bf 105}, 8797 (2001).

\bibitem{Beverly:02}
K.~Beverly, J.~Sampaio, and J.~Heath,
\newblock J. Phys. Chem. B {\bf 106}, 2131 (2002).

\bibitem{Remacle:98b}
F.~Remacle, C.~P. Collier, J.~R. Heath, and R.~D. Levine,
\newblock Chem. Phys. Lett. {\bf 291}, 453 (1998).

\bibitem{Baldea:2002}
I.~B\^aldea and L.~S. Cederbaum,
\newblock Phys. Rev. Lett. {\bf 89}, 133003 (2002).

\bibitem{Baldea:2007}
I.~B\^aldea and L.~S. Cederbaum,
\newblock Phys. Rev. B {\bf 75}, 125323 (2007).

\bibitem{Baldea:2008}
I.~B\^aldea and L.~S. Cederbaum,
\newblock Phys. Rev. B {\bf 77}, 165339 (2008).

\bibitem{Langreth:66}
D.~C. Langreth,
\newblock Phys. Rev. {\bf 150}, 516 (1966).

\bibitem{Shiba:75}
H.~Shiba,
\newblock Progr. Theor. Phys. {\bf 54}, 967 (1975).

\bibitem{Meir:92}
Y.~Meir and N.~S. Wingreen,
\newblock Phys. Rev. Lett. {\bf 68}, 2512 (1992).

\bibitem{Meir:94}
A.-P. Jauho, N.~S. Wingreen, and Y.~Meir,
\newblock Phys. Rev. B {\bf 50}, 5528 (1994).

\bibitem{Ng:88}
T.~K. Ng and P.~A. Lee,
\newblock Phys. Rev. Lett. {\bf 61}, 1768 (1988).

\bibitem{Izumida:01}
W.~Izumida, O.~Sakai, and S.~Suzuki,
\newblock J. Phys. Soc. Jpn. {\bf 70}, 1045 (2001).

\bibitem{Costi:01}
T.~A. Costi,
\newblock Phys. Rev. B {\bf 64}, 241310 (2001).

\bibitem{koeppel:84}
H.~K\"oppel, W.~Domcke, and L.~S. Cederbaum,
\newblock Adv. Chem. Phys. {\bf 57}, 59 (1984).

\bibitem{Baldea:97}
I.~B\^aldea, H.~K\"oppel, and L.~S. Cederbaum,
\newblock Phys. Rev. B {\bf 55}, 1481 (1997).

\bibitem{HHK:80}
D.~Bullet, R.~Haydock, V.~Heine, and M.~J. Kelly,
\newblock in {\em Solid State Physics}, edited by H.~Erhenreich, F.~Seitz, and
  D.~Turnbull, Academic, New York, 1980.

\bibitem{fulde:91}
P.~Fulde,
\newblock Electron correlations in molecules and solids,
\newblock in {\em Springer Series in Solid-State Sciences}, volume 100,
  Springer-Verlag (Berlin, Heidelberg, New York), 1991.

\bibitem{dagotto:94}
E.~Dagotto,
\newblock Rev. Mod. Phys. {\bf 66}, 763 (1994).

\bibitem{DelaneyGreer:06}
P.~Delaney and J.~C. Greer,
\newblock Proc. Roy. Soc. A {\bf 462}, 117 (2006).

\end{thebibliography}
\end{document}